\documentclass[11pt]{article}
\usepackage[sort&compress]{natbib}
\usepackage{fancyhdr}
\usepackage{isomath}
\usepackage{amsmath}
\usepackage{amsbsy}
\usepackage{amssymb}
\usepackage{amscd}
\usepackage{amsfonts}
\usepackage{graphicx,xcolor}
\usepackage{verbatim}
\usepackage{euscript}
\usepackage{alltt}
\usepackage{stmaryrd}
\usepackage{subfigure}
\usepackage{relsize}
\usepackage{cancel}
\usepackage{hyperref}
\usepackage[toc]{appendix}
\usepackage{longtable}
\hypersetup{
    colorlinks=true,
    allcolors = blue,
}
\DeclareGraphicsExtensions{.eps,.pdf}

\usepackage{nomencl}

\makenomenclature
\renewcommand{\nomgroup}[1]{%
    \ifthenelse{\equal{#1}{B}}{\item[\textbf{Other Variables}]}{}}
\RequirePackage{ifthen}

\usepackage[margin=1in]{geometry}
\DeclareMathAlphabet\mathbfcal{OMS}{cmsy}{b}{n}

\usepackage{amsmath}
\usepackage{amsbsy}
\usepackage{amssymb}
\usepackage{amscd}
\usepackage{amsfonts}

\newcommand{\R}{\mathbb R}

\newcommand{\bfb}{{\mathbf b}}

\newcommand{\bfe}{{\mathbf e}}

\newcommand{\bfm}{{\mathbf m}}

\newcommand{\bfs}{{\mathbf s}}

\newcommand{\bfv}{{\mathbf v}}

\newcommand{\bfx}{{\mathbf x}}

\newcommand{\bfA}{{\mathbf A}}
\newcommand{\bfB}{{\mathbf B}}

\newcommand{\bfF}{{\mathbf F}}

\newcommand{\bfH}{{\mathbf H}}
\newcommand{\bfI}{{\mathbf I}}

\newcommand{\bfK}{{\mathbf K}}
\newcommand{\bfL}{{\mathbf L}}

\newcommand{\bfT}{{\mathbf T}}

\newcommand{\bfV}{{\mathbf V}}
\newcommand{\bfW}{{\mathbf W}}
\newcommand{\bfX}{{\mathbf X}}

\newcommand{\eps}{{\epsilon}}

\newcommand{\beq}{\begin{equation}}
\newcommand{\eeq}{\end{equation}}
\newcommand{\beqs}{\begin{eqnarray}}
\newcommand{\eeqs}{\end{eqnarray}}
\newcommand{\beql}{\begin{equation} \label}

\newcommand{\calA}{{\mathcal A}}

\newcommand{\calD}{{\cal D}}
\newcommand{\calE}{{\cal E}}

\newcommand{\calL}{{\cal L}}
\newcommand{\calM}{{\cal M}}

\newcommand{\calP}{{\cal P}}

\newcommand{\calV}{{\cal V}}

\newcommand{\bfomega}{\boldsymbol{\omega}}

\newcommand{\bfLambda}{\boldsymbol{\Lambda}}
\newcommand{\bfGamma}{\boldsymbol{\Gamma}}
\newcommand{\bfOmega}{\boldsymbol{\Omega}}

\newcommand{\bfzero}{\mathbf{0}}

\newcommand{\curl}{\mathop{\rm curl}\nolimits}
\newcommand{\Div}{\mathop{\rm div}\nolimits}

\newenvironment{rcases}
{\left.\begin{aligned}}
{\end{aligned}\right\rbrace}

\newcommand{\p}{\partial}

\date{}
\begin{document}

\title{Dynamic micromagnetism a la Ericksen-Leslie, and the constrained polar continuum mechanics of hard magnetic soft materials}
%%allowing the Einstein-de Haas and Barnett effects\\
 %%Alternate title: On some links between the continuum mechanics of Ericksen-Leslie nematics, dynamic micromagnetism, and hard magnetic soft materials }

\author{Amit Acharya\thanks{Department of Civil \& Environmental Engineering, and Center for Nonlinear Analysis, Carnegie Mellon University, Pittsburgh, PA 15213, email: acharyaamit@cmu.edu.} \and Siladitya Pal\thanks{Department of Mechanical and Industrial Engineering, Indian Institute of Technology Roorkee, Roorkee, Uttarakhand, 247667 India, email:siladitya.pal@me.iitr.ac.in} }

\maketitle
$\qquad $ 
\begin{center}
\textit{Dedicated to a celebration of the career of Professor Robin Knops.\\}
\end{center}

\begin{abstract}
\noindent 
A model of dissipative micromagnetics coupled to (visco-)elasticity is explored, following the procedures of the Ericksen-Leslie theory of nematic liquid crystals allowing for angular momentum due to magnetization. An outcome is the Landau-Lifshitz-Gilbert theory coupled to material spin. A further power-less augmentation to the angular momentum of the theory with classical kinetic energy density is also considered, with a preliminary exploration of its potential in representing the Einstein-de Haas and Barnett effects within continuum mechanics. A treatment of the continuum mechanics of hard magnetic soft materials as a constrained polar material is presented. The models of \cite{desimone2002constrained} and \cite{zhao2019mechanics} are discussed as two different, namely energetically and kinematically, constrained models of magnetoelasticity encompassed within the overall framework.
\end{abstract}

\section{Introduction }
%%{ will probably change, given EdH-Barnett developments}

Up to the nature of the topological defects supported by them, there appear to be some similarities in the continuum representations of the magnetization in a ferromagnetic body and the director in nematic liquid crystals, especially when the latter is viewed through the lens of Ericksen-Leslie theory. We explore this connection in this work. Specifically, as a first objective, we derive a mechanical theory of motion coupled to dynamic micromagnetism starting from the balance laws of mass, momentum, angular momentum (accounting for the effects of magnetization in angular momentum of the body) and Maxwell's magnetostatics. A special feature of the derivation is to show that, methodologically, this follows exactly from the standard derivation of Ericksen-Leslie theory as developed in \cite{leslie1992continuum} (also see the textbook by \cite{stewart2007dynamic}), with a physically natural adjustment in the definition of the angular momentum of the body. The Landau-Lifschitz-Gilbert theory of dynamic micromagnetism (\cite{Landau1935,gilbert2004phenomenological, gilbert1956formulation}) emerges, with an added contribution arising from essentially frame-indifference of the possible dissipative coupling of the magnetization to the motion of the material. An `effective field' reflecting non-dissipative coupling to mechanical elasticity can be  read-off in the angular momentum balance\footnote{The Ericksen-Leslie theory in its original form does not account for positional solid elasticity (as appropriate for nematic liquid crystals). This theory has been extended to account for the behavior of liquid crystalline elastomers in \cite{anderson1999continuum}, including a treatment of nematic disclinations and dislocations in solids in \cite{acharya2014continuum}. We work with the development and notation of \cite{acharya2014continuum} here.} of the theory. For a restricted class of free energy density functions dependent only on the deformation gradient and the magnetization, our model provides a dynamic extension of the constrained theory of magnetostriction of \cite{desimone2002constrained, james1993theory} rooted in the balance principles and constitutive restrictions of continuum mechanics and Maxwell's magnetostatics.

     In addition, based on motivation from the well-accepted physical experiments pertaining to the Einstein-de Haas~\citep{einstein1915experimental} (EdH), Richardson \citep{richardson}, and Barnett \citep{barnett1915magnetization}  effects, we propose and explore a further natural, and experimentally testable, augmentation to the angular momentum density. Both changes to the angular momentum density arise in our model \emph{without any change to the classical expression for the kinetic energy density}. If on subsequent examination the second augmentation is not found to be in discord with other physical observations, this would allow the representation of the EdH and Barnett phenomena within the theory.

     Some of our results above can possibly be obtained within the very general framework of \cite{desimone1995inertial}, which starts from an additional postulated balance law of microscopic angular momentum\footnote{\cite{desimone1995inertial} is limited in its consideration of the dependence of the magnetic energy density on the magnetization gradient; it is not clear to us if that is merely for simplicity or is a technical requirement.}. The emphasis in our work is on somewhat more specificity, within a minimalistic approach w.r.t starting assumptions.

      In the process of these developments, we also consider the continuum mechanics of a polar material whose director spin is constrained to be the material spin, making contact with strain-gradient theories with couple stress of \cite{toupin1962elastic}\footnote{Our constrained theory is similar to, but does not belong to, the class discussed by \cite{toupin1962elastic} because the constrained director in our case cannot be expressed as a function of the instantaneous deformation gradient from some reference.} and \cite{mindlin1962effects}. Using this framework, as a second objective, we try to understand the theoretical structure of a demonstrably successful nonlinear model of micromagnetics for hard magnetic soft materials (\cite{zhao2019mechanics, wang2021evolutionary}) that entails the claim that ``rigid body rotation can change the free-energies of polar materials as well.'' We show here how the model employed by these authors arises as a special case of a frame-indifferent, \emph{constrained} theory for a polar material whose free-energy density is manifestly invariant to rigid body rotations (also see \cite{dorfmann2024hard} for related discussion). 
      
      Our considerations reveal an interesting feature about the modeling of materials in which deformation of solids is intended to be controlled by magnetic fields through the interaction with the magnetization. As we show, the model of constrained magnetoelasticity of \cite{desimone2002constrained} and that of hard magnetic soft materials \cite{zhao2019mechanics} (as interpreted by us) are two valid, but different, approaches, both of which can be used to model actuation of solids through an applied magnetic field. A detailed study of their range of applicability, particularly in relation to actuation in `hard' and `soft' solids, appears to be a useful topic for future study.

There is a vast literature on the formulation of various aspects of micromagnetism coupled to mechanical deformation, and it is beyond the scope of this paper to review it; we provide a limited review in the Appendix \ref{app:lit_rev}. Our aim in this contribution is to shed light on the two main goals elaborated above, with the hope that doing so illustrates one systematic way for deriving the continuum mechanics of polar materials, along with making a very direct analogy with the mechanics of liquid crystals.

An outline of the paper is as follows: Sec.~\ref{sec:notation} lists the primary notation employed. The theory is developed Sec.~\ref{sec:theory}. In Sec.~\ref{sec:add_rot_inertia} a power-less rotational inertia contribution is considered in continuum mechanics, retaining the classical definition of the kinetic energy density. Its implications are considered in the contexts of providing a  plausible model for the Einstein-de Haas and Barnett effects within continuum mechanics, as well as one for hard magnetic soft materials viewed as a constrained polar material. Sec.~\ref{sec:dual} contains some observations on treating the Second Law as a field equation and a 
solution/approximation scheme for the proposed models. Appendix \ref{app:lit_rev} contains a limited review of some of the literature related to our work.

\subsection{Notation}\label{sec:notation}
It suffices to consider a Rectangular Cartesian coordinate system for our work w.r.t whose basis $({\mathbf e}_i, i = 1,2,3)$ we express all tensor components. All partial derivatives refer to coordinates on the current configuration of the body, and $\nabla$ represents the gradient on the same configuration. A superposed dot on a symbol denotes the material time derivative. The inner product of two vectors is denoted as ${\mathbf a} \cdot {\mathbf b}=a_{i}b_{i}$. The trace inner product of two second order tensors is
denoted as ${\mathbf A}:{\mathbf B}=A_{ij}B_{ij}$. For a third order tensor ${\mathbf A}$ and a second order tensor ${\mathbf B}$, ${\mathbf A}:{\mathbf B} = {\mathbf A}_{ijk}B_{jk} {\mathbf e}_i$ is a vector; for ${\mathbf b}$ a vector ${\mathbf A}{\mathbf b} = {\mathbf A}_{ijk}b_{k}{\mathbf e}_i \otimes {\mathbf e}_j$ is a second order tensor.
The symbol ${\mathbf A}{\mathbf B}$ indicates the tensor multiplication of the second-order tensors ${\mathbf A}$ and ${\mathbf B}$. The notation $(\cdot)_{sym}$, $(\cdot)_{skw}$, and $(\cdot)_{dev}$ indicate
the symmetric, skew symmetric, and deviatoric parts, respectively, of the second order tensor $(\cdot)$. $\eps_{ijk}$ represent the permutation symbol.  

A list of some of the main mathematical symbols used is given below:
\begin{longtable}[h!]{l l}
 \endfirsthead
$[A]$ &  Physical dimension of quantity $A$\\
${\mathbf x}$ &  current position \\
${\mathbfcal M}$ & magnetization \\
${\mathcal M}_{s}$ & spontaneous magnetization ${\mathcal{M}_{s}} =\sqrt{{\mathbfcal M}\cdot {\mathbfcal M}}$\\
${\mathbf m}$ & unit magnetization ${\mathbf m}={\mathbfcal M}/{\mathcal M}_{s}$\\
${\mathbf F}$ & deformation gradient tensor\\
${\mathbf W}$ & {inverse elastic distortion tensor}; for the purposes of this paper ${\mathbf W} = {\mathbf F}^{-1}$  \\
${\mathbf v}$ & {material velocity} \\
${\boldsymbol \omega}$ & {magnetization spin (angular velocity) vector; ${\mathbf m} \times \dot {\mathbf m} =: {\boldsymbol \omega} \Longrightarrow \dot{\mathbf m} = {\boldsymbol \omega} \times {\mathbf m}$} \\
${\rho}$ & {mass density} \\
${\boldsymbol \Theta}$ & { gradient of magnetization spin vector ${\boldsymbol \Theta} = \nabla {\boldsymbol \omega}$}\\
${\mathbf T}$ & {Cauchy stress tensor}\\
${\boldsymbol \Lambda}$ & {couple stress tensor}\\
${\mathbf K}$& body moment per unit mass, including the torque from the stray/demagnetization\\
& field due to the magnetization of the body \\
${\mathbf b}$& {body force per unit mass}\\
$\psi$ & {free energy per unit mass} \\
$\mathbf H$ & {magnetic field}\\
${\mathbf H}_{a}$ & {applied magnetic field}\\
$\beta_m$ & viscosity associated with relative spin of magnetization w.r.t the material\\
${\mathbf X}$ & {(third order) alternating tensor $\bfX = \eps_{ijk} \bfe_i \otimes \bfe_j \otimes \bfe_k$}\\
${\bfI}$ & {the second order Identity tensor}\\
${\mathbf L}$ & {velocity gradient $\bfL := \nabla \bfv$} \\
${\mathbf D}$ & {rate of deformation tensor ${\mathbf D} = {\mathbf L}_{sym}$} \\
${\boldsymbol \Omega}$ & {material spin tensor ${\boldsymbol \Omega} = {\mathbf L}_{skw}$}\\
${\mathbf s}$ & {material spin vector ${\boldsymbol\Omega} {\mathbf b} =  {\mathbf s} \times {\mathbf b}$ for all vectors ${\mathbf b}$  }\\
${\boldsymbol \Gamma}$ & {negative of magnetization spin tensor (skew symmetric) $-{\boldsymbol \Gamma} {\mathbf b}$ =  $\boldsymbol{\omega} \times{\mathbf b}$ for all vectors ${\mathbf b}$ }
%%\end{tabular}
\end{longtable}

\section{Dynamic micromagnetism coupled to mechanical motion}\label{sec:theory}
We consider a body ${\mathcal V}(t)$ at time $t$ containing a fixed set of material particles with $\partial{\mathcal V}(t)$ as its boundary and ${\mathbf n}$ as its outward unit normal field. Conservation of mass requires that the material time derivative of total mass vanishes for all region ${\mathcal V}(t)$ and is expressed as  
\begin{align}
	\frac{d}{dt}\int_{{\mathcal V}(t)}\rho \, dv&=0, 
	 \label{eq1}
\end{align}
where $\rho$ is the mass density field.
Utilizing Reynolds' transport theorem, the equation 
of balance of mass (continuity) is obtained as 
\begin{align}\label{eq:bom}
    {\dot \rho}+\rho div {\mathbf v}&=0. 
\end{align}
The balance of linear momentum states that the change of linear momentum of any part is equal to the sum of surface and body forces acting on it and can be expressed as 
\begin{align}
	\frac{d}{dt}\int_{{\mathcal V}(t)}\rho {\mathbf v} dv&=\int_{ \partial {\mathcal V}(t)}{\mathbf T} {\mathbf n} da+\int_{{\mathcal V}(t)}\rho {\mathbf b} dv. 
\end{align}	
Applying Reynolds' transport theorem and balance of mass, we have
\begin{align}
    \rho \dot {\mathbf v}&=div {\mathbf T} +\rho {\mathbf b},
    \label{eq:balance_linear_momentum}
\end{align}
where the body force density $\bfb$ includes any body forces of magnetic origin.

To proceed further, we consider the Einstein-de Haas (EdH) effect~\citep{einstein1915experimental,richardson}, where the spin of electrons generates
a macroscopic torque on the crystal lattice. In the experiment, 
a cylinder made of ferromagnetic material was suspended inside a coil whose axis coincides with that of the cylinder. It was observed that on changing the flow direction of current within the solenoid
a measurable rotation of the cylinder about its axis was produced. %%In the classical interpretation, the net magnetization of the cylinder is assumed to be at most parallel to the axis of the cylinder at all times so that the magnetic field induced by the solenoid cannot apply a torque on the cylinder. 
The observed rotation is inferred to arise from a change in mechanical angular momentum induced by a change in the magnetization, in the absence of any mechanical torques being applied to the cylinder. 

As rough motivation, we consider a volume ${\mathcal V}$ which consists of $N_{n}$ number of nuclei and
of $N_{e}$ number of electron. Let the total angular momentum ${\mathbfcal L}$ $\left( [\calL] = \frac{Mass.Length^2}{Time}\right)$ be written as 
\begin{equation}
   {\mathbfcal L}= \sum_{i=1}^{N_{n}+N_{e}} m_{i} {\mathbf r}_{i}\times {\mathbf v}_{i}+\sum_{j=1}^{N_{e}} {\mathbfcal S}_{j}
    \label{eq:total_ang_momentum_discrete}
\end{equation}
where ${\mathbf r}_{i}$ and ${\mathbf v}_{i}$ span over the position 
and velocity of the nuclei and electrons. 
${\mathbfcal S}_{j}$ is the spin angular momentum of electron $j$ and expressed as 
\begin{equation}
    {\mathbfcal S}_{j}= \frac{2m_{e}}{{g}^{'}e} {\boldsymbol \mu}_{j}
    \label{eq:moment_of_electron}
\end{equation}
where ${\boldsymbol \mu}_{j}$ is the magnetic dipole moment $\left( [\mu_j] = \frac{Charge . Length^2}{Time}\right)$ of the electron $j$
and $e$ is the $(-)$ve charge of an electron $\left([e] = Charge\right)$. $m_{e}$ is the mass of an electron 
and $g^{'}$ is the Lande g-factor. Substituting this expression in \eqref{eq:moment_of_electron}, the total angular momentum \eqref{eq:total_ang_momentum_discrete} is expressed as 
\begin{equation}
{\mathbfcal L}= \sum_{i=1}^{N_{n}+N_{e}} m_{i} {\mathbf r}_{i}\times {\mathbf v}_{i}+\sum_{j=1}^{N_{e}} -\frac{2m_{e}}{{g'}|e|} {\boldsymbol \mu}_{j}
\end{equation}
 Now for externally applied torque ${\mathbfcal T}$, the conservation of angular momentum states that $\dot{\mathbfcal L}={\mathbfcal T}$, i.e.,
\begin{equation}\label{eq:sign_g}
    \frac{d}{dt}\bigg[\sum_{i=1}^{N_{n}+N_{e}} m_{i} {\mathbf r}_{i}\times {\mathbf v}_{i}+\gamma^{-1}\sum_{j=1}^{N_{e}} {\boldsymbol \mu}_{j}\bigg]={\mathbfcal T}
\end{equation}
Furthermore, the gyromagentic ratio is defined
as $\gamma= g'e/(2m_{e})$ $\left([\gamma] = \frac{Charge}{Mass} \right)$. The magnetization ${\mathbfcal M}(\mathbf x,t)$ is defined as net dipole moment per unit mass $\left([\calM] = \frac{Charge.Length^2}{Mass.Time}\right)$ and we assume the following relationship: 
\begin{equation}
    \int_{{\mathcal V}(t)}{\mathbfcal M}({\mathbf x},t)\rho({\mathbf x},t)dv=\sum_{j=1}^{N_{e}} {\boldsymbol \mu}_{j}.  
\end{equation}
Motivated by the above heuristics, we assume a continuum form of balance
of angular momentum as
\begin{equation}
    \frac{d}{dt}\int_{{\mathcal V}(t)}\rho[{\mathbf x }\times{\mathbf v}+\gamma^{-1}{\mathbfcal M}] \, dv= {\mathbfcal T}.
    \label{eq:continuum_E_dh}
\end{equation}
This equation would qualitatively seem to explain the rotation of the solid body upon magnetization i.e., Einstein-de Haas effect when there is no external moment (${\mathbfcal T} = {\mathbf 0}$) and the Barnett effect \citep{barnett1915magnetization} where magnetic moment 
is generated through mechanical rotation, but as we will see, a closer examination will require more. In an earlier work, ~\citet{alblas1968cosserat} (and several more recent works) included electronic spin
 in the kinetic energy density of the Lagrangian  describing the deformation of a ferromagnetic material and used a variational approach to obtain a statement of angular momentum balance which contains spin angular momentum. 
 
 In contrast, in our formalism the total kinetic energy is not affected by electronic spin, a fact we interpret as in keeping with the non-material nature of electronic spin. 
 
 Upon consideration of all torques acting on a part through its surface and volume, balance of angular momentum can be stated as 
\begin{align}	
	\frac{d}{dt}\int_{{\mathcal V}(t)}\rho \big({\mathbf x }\times{\mathbf v}+\gamma^{-1}{\mathbfcal M} \big)dv&
	=\int_{\partial {\mathcal V}(t)} \big({\mathbf x }\times{\mathbf T} +{\boldsymbol \Lambda}\big) {\mathbf n} da
	+\int_{{\mathcal V}(t)} \rho\big({\mathbf x }\times{\mathbf b} +{\mathbf K}\big) dv.
	\label{eq:balance_angular_momentum}
\end{align}

Finally, using the Reynolds' transport theorem, the local form of balance of angular momentum can be identified as  
\begin{align}
	       \gamma^{-1} \rho \dot {\mathbfcal M}&=div {\boldsymbol \Lambda} -{\mathbf X}:{\mathbf T}+\rho {\mathbf K}.
	       \label{eq:angular_momentun}
\end{align}
Here we define the unit magnetization ${\mathbf m}={\mathbfcal M}/{\mathcal M}_{s}$, where ${\mathcal M}_{s}$ is the spontaneous (saturation) magnetization. \emph{In the following, we will, with some abuse of notation, often refer to the unit magnetization as the magnetization.}
As  the processes considered will be  well below the Curie temperature, changes in ${\mathcal M}$ will be assume to arise only from the direction of the unit magnetization, with the magnitude of ${\mathcal M}$ remaining constant:
\begin{align}
    {\mathbf m}\cdot {\mathbf m}-1=0, \:\:\textrm{and}\:\: |{\mathbfcal M}|={\mathcal M}_{s}.
\end{align}
 Therefore, the balance law can be presented in terms of the unit magnetization as 
\begin{align}
	       \frac{\rho {\mathcal M}_{s}}{\gamma} \dot  {\mathbf m}&=div {\boldsymbol \Lambda} -{\mathbf X}:{\mathbf T}+\rho {\mathbf K}.
	       \label{eq:angular_momentum_0}
\end{align}
The body torque per unit mass, $\bfK$ $\left([K] = \frac{Length^2}{Time^2} \right)$, consists of the sum of externally applied mechanical body torque densities and  those of magnetic origin. The latter is assumed here to be given by \cite[Secs.~5.6-5.7]{jacksonCED}
\[
\bfK_m := \calM_s \bfm \times \bfB,
\]
where $\bfB$ is the magnetic induction $\left([B] = \frac{Mass}{Charge.Time} \right)$. We assume that the magnetic field, $\bfH$ $\left([H] = \frac{Charge}{Time.Length} \right)$, is the sum of an externally applied magnetic field $\bfH_a$ (which satisfies $\Div \bfH_a = 0$ and $\curl \bfH_a = \bfzero$ on $\calV(t)$ for all $t$), and the `stray' or demagnetization field $\bfH_s$. 
%%The total torque of magnetic origin is given by$$\bfK_m = \bfm \times \bfH; \qquad \qquad \bfH := \bfH_s + \bfH_a.$$
In the absence of (macroscopic) electric currents and charges and on timescales where it suffices to consider magnetic effects to be quasi-static, the demagnetization field $\bfH_s$ satisfies the equations of magnetostatics in the presence of matter at macroscopic scales \cite[Sec.~5.8; a $\rho$ appears in \eqref{eq:mag_stat}$_2$ since our magnetization is defined a per unit mass measure]{jacksonCED} given by
\begin{equation}\label{eq:mag_stat}
%%\begin{aligned}
\begin{rcases}
    \curl \bfH_s &= \bfzero \\
    \Div  ( \bfH_s +  {\rho}\calM_s \bfm) & = 0 
    \end{rcases} \qquad \mbox{ on } \mathcal{E}^3,
%%\end{aligned}
\end{equation}
where $\calE^3$ represents all of 3-d Euclidean space whose representation in the chosen rectangular Cartesian coordinates is $\R^3$. The magnetostatics system involves a (non-local) coupling of the magnetization to the stray/demagnetization field $\bfH_s$. Since $\R^3$ is simply connected, a potential $\phi$ is often introduced to write $\bfH_s = - \nabla \phi$ and \eqref{eq:mag_stat}$_{2}$ solved in the sense of distributions (or in weak form) for $\phi$ with $\bfm$ extended outside the material body in the current configuration by $\bfzero$.

The magnetic induction $\bfB$ is given by
\[
    \mu_0(\bfH_a + \bfH_s +  {\rho}\calM_s \bfm)  = \bfB
\]
\cite[Eqn.~5.81]{jacksonCED}, where $\mu_0$ is the magnetic permeability of vacuum $\left([\mu_0] = \frac{Mass.Length}{Charge^2} \right)$. This leads to the body torque density of magnetic origin to be given by
\begin{equation}\label{eq:K_m}
    \bfK_m = (\mu_0 \calM_s) \,\bfm \times (\bfH_a + \bfH_s).
\end{equation}
\emph{Equations \eqref{eq:bom}, \eqref{eq:balance_linear_momentum}, \eqref{eq:angular_momentum_0}, \eqref{eq:mag_stat} form the basic balance laws of concern in this work.}

In Secs.~\ref{sec:theory}-\ref{sec:add_rot_inertia}, we will consider the expression for the external power supplied to the body as given by \eqref{eq:pwr_exp} below. 
In Sec.~\ref{sec:dual} we will consider the possibility of  operating with the external power to be defined as
the replacement of $\bfK$ in \eqref{eq:pwr_exp} by $\bfK - (\mu_0 \calM_s) \bfm \times \bfH_s$.

We do not dwell on the philosophical implications of making a choice on the form of the external power supplied to the body in various models of continuum thermomechanics (coupled to magnetostatics here), noting that flexible frameworks allowing for different modeling choices have been invoked in the continuum mechanics literature, see, e.g., \cite[Chs.~84, 92]{gurtin2010mechanics}, \cite{desimone1995inertial},  and \cite[Ch.~15]{kovetz2000electromagnetic}.

\begin{comment}
===========================================
The body moment ${\mathbf K}$ when magnetization is subjected to an external magnetic field ${\mathbf H}$ can take a form~\citep{stewart2007dynamic} 
\begin{equation}
 \rho {\mathbf K}={\mathbf m}\times {\mathbf G},\quad\quad {\mathbf G}=\chi_{a}({\mathbf m}\cdot{\mathbf H}){\mathbf H}
\end{equation}
where, $\chi_{a}$ is the magnetic anisotropy. Similarly, the body force ${\mathbf b}$ in Equation~(\ref{eq:balance_linear_momentum}) is also found as 
\begin{equation}
   \rho {\mathbf b}=({\mathbf m}\cdot \nabla) {\mathbf H}
\end{equation}
Utilizing the Gauss's law (\eqref{eq:mag_induc}) and the Ampere's law (\eqref{eq:field}), the magnetic body force can be expressed as the divergence of a second order tensor~\citep{brown1966magnetoelastic}, the so called Maxwell's stress tensor as 
\begin{equation}
   \rho{\mathbf b}=div {\mathbf T}^{m}
\end{equation}
with ${\mathbf T}^{m}$=${\mathbf B}\otimes{\mathbf H}-\mu_{0}({\mathbf H}\cdot{\mathbf H}){\mathbf I}$.
%=================================================================
\end{comment}

The external power supplied at any given time can be expressed as 
\begin{subequations}\label{eq:thm_power_expend}
\allowdisplaybreaks
\begin{align}
	{\mathcal P}_{ext}(t)&=\int_{{\mathcal V}(t)}\rho {\mathbf b}\cdot {\mathbf v} dv+\int_{\partial {\mathcal V}(t)}{\mathbf T} {\mathbf n}\cdot {\mathbf v}da+
	\int_{\partial {\mathcal V}(t)}{\boldsymbol  \Lambda} {\mathbf n}\cdot {\boldsymbol \omega}da+\int_{{\mathcal V}(t)}\rho {\mathbf K}\cdot {\boldsymbol \omega}dv \label{eq:pwr_exp}\\
	&=\int_{{\mathcal V}(t)}\big(\rho {\mathbf v}\cdot \dot{\mathbf v}
	+ 
	{\frac{\rho {\mathcal M}_{s}}{\gamma}\dot{\mathbf m}\cdot {\boldsymbol \omega}}  \big) 
	dv
	+\int_{{\mathcal V}(t)}\big({\mathbf T}:{\mathbf L}+{\boldsymbol \Lambda}:{\boldsymbol \Theta}+{\mathbf T}:{\boldsymbol \Gamma}\big) dv \notag \\
		&=\int_{{\mathcal V}(t)}\rho {\mathbf v}\cdot \dot{\mathbf v} dv
	+\int_{{\mathcal V}(t)}\big({\mathbf T}:{\mathbf L}+{\boldsymbol \Lambda}:{\boldsymbol \Theta}+{\mathbf T}:{\boldsymbol \Gamma}\big) dv \label{eq:red_pwr_exp}
\end{align}
\end{subequations}
since 
\[
{\dot {\mathbf m}}\cdot {\boldsymbol \omega} = ({\boldsymbol \omega} \times {\mathbf m}) \cdot {\boldsymbol \omega} = 0.
\]
The kinetic energy and free energy of the body are defined as 
\begin{align}
	{\mathcal K}&=\int_{{\mathcal V}(t)}\frac{1}{2}\rho {\mathbf v}\cdot {\mathbf v} dv \:\:\: \textrm{and} \:\:\:
	{\mathcal F}=\int_{{\mathcal V}(t)} \rho \psi dv,
\end{align}
respectively. Using Reynold's transport theorem, we obtain the mechanical dissipation 
\begin{align}\label{eq:diss}
	\mathcal D:={\mathcal P}_{ext}-\frac{d}{dt}\bigg[{\mathcal K}+{\mathcal F}\bigg]=\int_{{\mathcal V}(t)}\big({\mathbf T}:{\mathbf L}+{\boldsymbol \Lambda}:{\boldsymbol  \Theta}+{\mathbf T}:{\boldsymbol \Gamma}-\rho \dot \psi\big) dv.
\end{align}

\subsection{Frame indifference and Ericksen's identity for magnetomechanics}\label{sec:frame_ind}
We consider invariance under superposed rigid body motions of the free energy density function $\psi$, where $\psi$ depends on 
${\mathbf m}$, $\nabla {\mathbf m} $ and ${\mathbf W}$ \citep{ericksen1961conservation,stewart2007dynamic,anderson1999continuum}. Invariance under superposed rigid motions requires that $\psi$, being a physical scalar density, must satisfy
\begin{equation}
	\psi({\mathbf m},\nabla {\mathbf m}, {\mathbf W})=\psi({\mathbf Q}{\mathbf m},{\mathbf Q}\nabla {\mathbf m}{\mathbf Q}^{T},
	{\mathbf W}{\mathbf Q}^{T})
\end{equation}
for all proper orthogonal second order tensors ${\mathbf Q}$, and for all elements ${\mathbf m}$, $\nabla {\mathbf m} $, ${\mathbf W}$ in 
the domain of the function $\psi$. Considering ${\mathbf S}$ to be an arbitrary fixed skew tensor, we parameterize the rotation ${\mathbf Q}(s)$ for 
all scalar values of $s$ such that
\begin{equation}
	{\mathbf Q}(0)={\mathbf I}; \quad \frac{d{\mathbf Q}}{ds}(0)={\mathbf S}=-\frac{d{\mathbf Q}^{T}}{ds}(0).
\end{equation}
Differentiating the above equation and then evaluating it at $s = 0$ gives
\begin{equation}
	\bigg[\frac{\partial \psi }{\partial {\mathbf m} }\otimes {\mathbf m}+ 
	\frac{\partial \psi }{\partial \nabla {\mathbf m} } \nabla {\mathbf m}^{T} 
	-\nabla {\mathbf m}^{T}\frac{\partial \psi }{\partial \nabla {\mathbf m} } 
	-{\mathbf W}^{T}\frac{\partial \psi }{\partial {\mathbf W} }  \bigg]:{\mathbf S}=0.
\end{equation}
Due to the arbitrariness of ${\mathbf S}$, Ericksen's identity for the magneto-elastic material can be obtained as 
\begin{equation}
	\bigg(\frac{\partial \psi }{\partial {\mathbf m} }\otimes {\mathbf m}\bigg)_{skw}=
	-\bigg(\frac{\partial \psi }{\partial \nabla {\mathbf m} } \nabla {\mathbf m}^{T} 
	-\nabla {\mathbf m}^{T}\frac{\partial \psi }{\partial \nabla {\mathbf m} }\bigg)_{skw}
	+ \bigg({\mathbf W}^{T}\frac{\partial \psi }{\partial {\mathbf W} }  \bigg)_{skw}.
	\label{eq:Ericksen_identity}
\end{equation}

\subsection{Constitutive relations and driving forces}\label{sec:const_eqn}
The material time derivative of $\psi$ is
\begin{align}
	\dot {\psi}=\frac{\partial \psi }{\partial {\mathbf m} }\cdot \dot{\mathbf m}
	+\frac{\partial \psi }{\partial \nabla {\mathbf m} }: \dot {\overline{\nabla {\mathbf m}}}
	+ \frac{\partial \psi }{\partial {\mathbf W} }: \dot{\mathbf W}.
	\label{eq:free_energy_derivative}
\end{align}
The material time derivative of the gradient of magnetization director is given as 
\begin{equation}
	\dot {\overline{\nabla {\mathbf m}}}=\mathrm{grad}\; \dot{\mathbf m}-{\nabla {\mathbf m}}{\mathbf L}.
	\label{eq:mag_grad_derivative}
\end{equation}
The first term of the time derivative of the free energy \eqref{eq:free_energy_derivative} can be written as (using $\dot{\bfm} = - \bfGamma \bfm$)
\begin{equation}
\frac{\partial \psi }{\partial {\mathbf m} }\cdot \dot{\mathbf m}=-\frac{\partial \psi }{\partial {\mathbf m} }\otimes {\mathbf m}:{\boldsymbol \Gamma}.
\end{equation}
Using (\ref{eq:mag_grad_derivative}), the second term of the 
time derivative of free energy(\eqref{eq:free_energy_derivative} can be simplified as 
\begin{equation}
	\frac{\partial \psi }{\partial \nabla {\mathbf m} }: \dot {\overline{\nabla {\mathbf m}}}
	= \bigg[{\mathbf X}:\left({\mathbf m}\otimes\frac{\partial \psi }{\partial \nabla {\mathbf m} }\right)\bigg]:{\boldsymbol  \Theta}
	- \bigg[\frac{\partial \psi }{\partial \nabla {\mathbf m} } {\nabla {\mathbf m} }^{T}\bigg]:{\boldsymbol \Gamma}
	- {\nabla {\mathbf m} }^{T}\frac{\partial \psi }{\partial \nabla {\mathbf m} }:{\mathbf L},
\end{equation}
where, for $\bfA$ a third order tensor, $\bfB$ a second order tensor, and $\bfb$ a vector, $\bfA : (\bfb \otimes \bfB)= A_{ijk} b_j B_{kr} \bfe_i \otimes \bfe_r$.

For $\bfW = \bfF^{-1}$,
\begin{equation}
\dot {\mathbf W}=-{\mathbf W}{\mathbf L} \Leftrightarrow  \dot{\mathbf F}{\mathbf F}^{-1}={\mathbf L},
\label{eq:compatible}
\end{equation}
with $\mathbf{F}$ being the deformation gradient w.r.t a fixed reference\footnote{ The presence of defects of compatibility in the elastic distortion is characterized by $\bfW \neq \bfF^{-1}$, $\curl \bfW \neq \bfzero$, which is the fundamental reason for plasticity of solids. The interaction of magnetization and plasticity is a subject of scientific interest \citep{alshits2017experimental}, studied by eminent figures in both subjects like W.F. Brown Jr., and A.~Seeger \cite[Sec.~1]{seeger1966effect}. An approach to studying magnetoplasticity, a subject of importance  in technology as well, see e.g., \cite{guo2022magneto}, is provided by a natural extension of the developments of the present work by the augmentation of \eqref{eq:compatible} by the fundamental kinematic statement $\dot {\mathbf W}+{\mathbf W}{\mathbf L}={\boldsymbol \alpha} \times {\mathbf V}$ \citep{acharya2004constitutive}, where dislocations are viewed through their density, physically contemplated as an areal density of lines
carrying Burgers vector. In this augmented statement, ${\boldsymbol \alpha}$ and ${\mathbf V}$ are  the dislocation density tensor and velocity field of the dislocation lines, respectively, and  ${\boldsymbol \alpha}\times {\mathbf V}$ is the local plastic distortion rate produced by
motion of dislocations. Thermodynamic restrictions for $\bfV$ have been derived \citep{acharya2004constitutive,acharya2011microcanonical} and used, and would fit into a natural extension of the current framework as can be seen from the development in \cite{acharya2014continuum}.}.
Utilizing \eqref{eq:compatible}, the third term of \eqref{eq:free_energy_derivative} can be written as 
\begin{equation}
    \frac{\partial \psi }{\partial {\mathbf W} }: \dot{\mathbf W}=-{\mathbf W}^{T}\frac{\partial \psi }{\partial {\mathbf W} }: {\mathbf L}.
\end{equation}
Restating the dissipation 
\begin{equation}
\mathcal D=\int_{{\mathcal V}(t)}\big({\mathbf T}:{\mathbf L}+{\boldsymbol \Lambda}:{\boldsymbol \Theta}+{\mathbf T}:{\boldsymbol \Gamma}-\rho {\dot \psi}\big) dv
\end{equation}
and inserting the components of $\dot \psi$ in the above dissipation inequality and rearranging the terms for ${\mathbf L}$,
${\boldsymbol \Gamma}$, ${\boldsymbol \Theta}$, we obtain
\begin{align}
	\bigg[{\mathbf T}+\rho{\nabla {\mathbf m} }^{T}\frac{\partial \psi }{\partial \nabla {\mathbf m} }+
	\rho{\mathbf W}^{T}\frac{\partial \psi }{\partial {\mathbf W} }\bigg]:{\mathbf L}
	+\bigg[{\mathbf T}+\rho\frac{\partial \psi }{\partial {\mathbf m} }\otimes {\mathbf m}+\rho\frac{\partial \psi }{\partial \nabla {\mathbf m} } {\nabla {\mathbf m} }^{T}\bigg]_{skw}:{\boldsymbol \Gamma} \nonumber \\
	+\bigg[{\boldsymbol \Lambda} - \rho{\mathbf X}:{\mathbf m}\otimes\frac{\partial \psi }{\partial \nabla {\mathbf m} }\bigg]:{\boldsymbol \Theta}
    \label{eq:31}
\end{align}
as the integrand of $\calD$ in \eqref{eq:31}. Motivated by the structure of the dissipation and what is known about the constitutive equations of complex fluids and viscoelastic solids, we consider the total stress to be of the form ${\mathbf T} = {\mathbf T}^{e} + {\mathbf T}^{d}$, the sum of an equilibrium and a dissipative part, with
\begin{equation}
	{\mathbf T}^{e}=-\rho{\nabla {\mathbf m} }^{T}\frac{\partial \psi }{\partial \nabla {\mathbf m} }-
	\rho{\mathbf W}^{T}\frac{\partial \psi }{\partial {\mathbf W} }.
	\label{eq:total_stress}
\end{equation}
Additionally the couple stress is assumed to be
\begin{equation}\label{eq:couple_stress}
{\boldsymbol \Lambda} = \rho{\mathbf X}:{\mathbf m}\otimes\frac{\partial \psi }{\partial \nabla {\mathbf m} }.
\end{equation}
The residual dissipation can then be expressed as
\begin{equation}
	\mathcal{D}=	{\mathbf T}^{d}_{sym}:{\mathbf D}+{\mathbf T}^{d}_{skw}:({\boldsymbol \Omega}-{\boldsymbol \Gamma}^{T})+\bigg[{\mathbf T}^{e}+\rho\frac{\partial \psi }{\partial {\mathbf m} }\otimes {\mathbf m}+\rho\frac{\partial \psi }{\partial \nabla {\mathbf m} } {\nabla {\mathbf m} }^{T}\bigg]_{skw}:{\boldsymbol \Gamma}\ge0.
\end{equation}
Utilizing the constitutive equation for the equilibrium stress as given in (\ref{eq:total_stress}) and Ericksen's identity (\ref{eq:Ericksen_identity}), the dissipation  inequality can be further simplified to
\begin{equation}
\mathcal{D}=	{\mathbf T}^{d}_{sym}:{\mathbf D}+{\mathbf T}^{d}_{skw}:({\boldsymbol \Omega}-{\boldsymbol \Gamma}^{T})\ge 0.
\end{equation}
In order to consider the viscous dissipation associated with the 
magnetization, we follow 
nematic liquid crystal theory \citep{stewart2007dynamic}, replacing the nematic director by the unit magnetization vector. Then, the form of the viscous stress is given by
\begin{align}
	{\mathbf T}^{d}&=\alpha_{1}({\mathbf m }\cdot {\mathbf D}{\mathbf m }){\mathbf m}\otimes{\mathbf m}
	+\alpha_{2}{\mathring {\mathbf m}}\otimes {\mathbf m } +\alpha_{3} {\mathbf m }\otimes {\mathring {\mathbf m}} \nonumber \\
	&+\alpha_{4}{\mathbf D}+\alpha_{5}({\mathbf D}{\mathbf m})\otimes{\mathbf m}+\alpha_{6} {\mathbf m}\otimes{\mathbf D}{\mathbf m}
	\label{eq:viscous_stress}
\end{align}
where $\mathring{\mathbf m} = \dot {\mathbf m}-{\boldsymbol \Omega}{\mathbf m}$ is the co-rotational time flux of the unit magnetization ${\mathbf m}$, related to its relative velocity w.r.t that produced by material spin. The $\alpha$ are the viscosity coefficients.
We note that the above form of the viscous stress involves the assumption of reflection symmetry of the liquid crystal director resulting in an isotropic ${\mathbf T}^{d}$. In contrast, due to the lack of such symmetry for a magnetization director,  a more generic form of the viscous stress (anisotropic) can be considered. However, for the sake of simplicity, we retain the isotropic form 
of the viscous stress tensor\footnote{Since a primary goal of our work is to demonstrate that the formalism directly leads to the emergence of a coupling of the magnetization to the material spin through the balance of angular momentum, a feature that will not be destroyed with the more realistic modeling of the viscous stress, we retain this simplicity here, leaving the full generality for future work. The necessary modifications can be developed by following the works of \citep{leslie1979thermodynamics, cowin1974theory}.}.
Inserting this into the expression for the dissipation, we have 
\begin{align}
	\mathcal{D}&=
	\alpha_{1} ({\mathbf m }\cdot {\mathbf D}{\mathbf m })^{2}
	+(\alpha_{2}+\alpha_{3}+\alpha_{6}-\alpha_{5}) \mathring {\mathbf m }\cdot {\mathbf D}{\mathbf m } \\
	&+\alpha_{4}{\mathbf D}:{\mathbf D}+(\alpha_{5}+\alpha_{6})   {\mathbf m}\cdot{\mathbf D}{\mathbf D}{\mathbf m}
	+(\alpha_{3}-\alpha_{2}){\mathring {\mathbf m}}\cdot {\mathring {\mathbf m}}.
\end{align}
The viscous stress can have several possible dependencies on ${\mathbf m}$ and ${\mathbf D}$ \eqref{eq:viscous_stress}. Considering the simplest alternative for our purposes to show a coupling between the magnetization dynamics and mechanical motion, we consider $\alpha_{1} = \alpha_{5} = \alpha_{6} = 0$ to obtain  a viscous stress given by
\begin{align}
	{\mathbf T}^{d}&=\alpha_{2}{\mathring {\mathbf m}}\otimes {\mathbf m } +\alpha_{3} {\mathbf m }\otimes {\mathring {\mathbf m}}+\alpha_{4}{\mathbf D}.
	\label{eq:dissipative_stress_tensor}
\end{align}
The first and second terms in \eqref{eq:dissipative_stress_tensor} represent the viscous stress contributions associated with the magnetization dynamics
while the third term indicates a part of the standard viscous stress of a Newtonian fluid or linear viscoelasticity for solids. In the following section, we will derive a statement for the balance of angular momentum which relates the skew symmetric part of the Cauchy stress and couple stress tensors to the dynamics of the magnetization.   
\subsection{Derivation of magnetization dynamics}
To derive the magnetization dynamics, we substitute the expression for the Cauchy stress tensor ${\mathbf T}$
and the couple stress tensor ${\boldsymbol \Lambda}$ in the expression of balance of angular momentum as given in (\ref{eq:angular_momentun}). Accordingly,
balance of angular momentum can be expressed as 
\begin{equation} \label{eq:angmom}
	\gamma^{-1} \rho {\mathcal M}_{s}\dot {\mathbf m} ={\mathbf X}: \Bigg[\rho\textrm{div}\bigg( {\mathbf m}\otimes\frac{\partial \psi }{\partial \nabla {\mathbf m} } \bigg) 
	+\rho{\nabla {\mathbf m} }^{T}\frac{\partial \psi }{\partial \nabla {\mathbf m} }+
	\rho{\mathbf W}^{T}\frac{\partial \psi }{\partial {\mathbf W} } -{\mathbf T}^{d}\Bigg]
	+\rho {\mathbf K}.
\end{equation}
Substituting Ericksen's identity \eqref{eq:Ericksen_identity} and the dissipative stress tensor \eqref{eq:dissipative_stress_tensor}, the above equation can be 
simplified to 
\begin{equation}\label{eq:llg_simplest}
\gamma^{-1}\rho {\mathcal M}_{s} \dot {\mathbf m}={\mathbf m}\times \Bigg[\rho\textrm{div}\bigg( \frac{\partial \psi }{\partial \nabla {\mathbf m} } \bigg) 
	 -\rho\frac{\partial \psi }{\partial  {\mathbf m} }-\beta_m {\mathring {\mathbf m}} \Bigg]
	+\rho {\mathbf K}; \qquad \qquad \beta_m = \alpha_{3}-\alpha_{2},
\end{equation}
where $\beta_m$ is the viscosity associated with the motion of the unit magnetization \emph{relative} to its motion due to material spin.

We now assume that the operative body torque density $\bfK$ is entirely composed of that arising from the magnetic induction given by \eqref{eq:K_m}:
$${\mathbf K}= \mu_0 \calM_s {\mathbf m}\times {\mathbf H} = { \mu_0} \calM_s {\mathbf m}\times ({\mathbf H_a} + \bfH_s).$$ %%, where ${\mathbf H}_{a}$ is the externally applied field.
On further
replacing the corotational flux of magnetization, \eqref{eq:llg_simplest} can be recast as
\begin{equation}
	\rho {\dot {\mathbf m}}=\gamma{\mathbf m}\times \frac{\rho}{{\mathcal M}_{s}} \Bigg[\textrm{div}\bigg( \frac{\partial \psi }{\partial \nabla {\mathbf m} } \bigg) 
	 -\frac{\partial \psi }{\partial  {\mathbf m} }\Bigg] +\frac{\gamma \rho}{{\mathcal M}_{s}} {\mathbf m}\times { \mu_0}\calM_s {\mathbf H} -\frac{\gamma \beta_m}{{\mathcal M}_{s}}{\mathbf m}\times \bigg[\dot {\mathbf m}- {\boldsymbol \Omega}{\mathbf m}\bigg].
\end{equation}
	%+\gamma \rho {\mathbf K}-\gamma {\beta_m{\mathbf m}\times \bigg[\dot {\mathbf m}- {\boldsymbol \Omega}{\mathbf m}\bigg]
This equation clearly indicates the contributions of the couple stress tensor, the body couple and the material spin-dependent rotational viscous stress towards the dynamics of magnetization. We further simplify the above equation by defining an effective magnetic field ${\mathbf H}_{eff}$ as 
\begin{equation}\label{eq:_Heff}
	{\mathbf H}_{eff}=\frac{1}{{\mathcal M}_{s}} \Bigg[\textrm{div}\bigg( \frac{\partial \psi }{\partial \nabla {\mathbf m} } \bigg)- \frac{\partial \psi }{\partial  {\mathbf m} }+ { \mu_0}\calM_s{\mathbf H}\Bigg] 
	 \end{equation}
$\left([H_{eff}] = \frac{Mass}{Time . Charge}\right)$. Thus, the magnetization dynamics can be expressed as 
\begin{equation}
	{\dot {\mathbf m}}=\gamma {\mathbf m}\times \bigg[{\mathbf H}_{eff}   - \eta \dot {\mathbf m}+\eta {\boldsymbol \Omega}{\mathbf m}\bigg]; \qquad \qquad \eta := \frac{\beta_m}{\rho \mathcal{M}_s}
	\label{eq:llg_direct}
\end{equation}
$\left([\eta] =\frac{Mass}{Charge}\right)$. For $\bfOmega = \bfzero$ (i.e., absence of material motion), we refer to this form of the balance of angular momentum as the `Gilbert form of the LLG equation'.

 The material time derivative of the magnetization, $\dot{\bfm}$, in \eqref{eq:llg_direct} can be directly solved for as
 \begin{equation*}
 \begin{aligned}
 \dot{\bfm} & = \mathbb{M}^{-1}  \left(\gamma \bfm \times \left(  \bfH_{eff} + \eta \bfOmega \bfm \right) \right),\\
 \mathbb{M} & := \bfI - \gamma \eta \bfX \bfm,\\
 \mathbb{M}^{-1} & = \bfI + \frac{\gamma \eta}{1 + \gamma^2 \eta^2} \bfX \bfm \ + \  \frac{\gamma^2 \eta^2}{1 + \gamma^2 \eta^2}\left(\bfX \bfm \right)^2
 \end{aligned}
 \end{equation*}
(with $\bfX \bfm = \eps_{ijk} m_k \,\bfe_i \otimes \bfe_j$, and $\mathbb{M}$ always invertible). Alternatively, the same result is obtained by substituting the expression for 
$\dot{\bf m}$ from \eqref{eq:llg_direct} into the occurrence of the same term in the r.h.s. of the equation itself, i.e., \eqref{eq:llg_direct}, and noting that $\bfm \times (\bfm \times \dot{\bfm}) = - \dot{\bfm}$:
\begin{align} 	\label{eq:llg_finite}
    	{\dot {\mathbf m}}=&\frac{\gamma}{1+\gamma^2\eta^2}{\mathbf m}\times \bigg[{{\mathbf H}_{eff}+\eta {\boldsymbol \Omega}{\mathbf m}}\bigg] - \frac{\gamma^2\eta}{1+\gamma^2\eta^2}{\mathbf m}\times\bigg({\mathbf m}\times \bigg[{{\mathbf H}_{eff}}+\eta {\boldsymbol \Omega}{\mathbf m}\bigg]\bigg).
    \end{align}
On recalling our definition $\gamma= \frac{g'e}{2m_e}$ (see the discussion surrounding \eqref{eq:moment_of_electron}-\eqref{eq:sign_g}) to introduce 
\[
\bar{\gamma} := - \gamma \mbox{ and } \lambda : =   \xi \bar{\gamma}, \mbox{ with the non-dimensional parameter } \  \xi := \bar{\gamma} \eta,
\]
\eqref{eq:llg_finite} can be rewritten as 
\begin{align}\label{eq:llg_simplified}
	\left( 1 + \xi^2\right){\dot {\mathbf m}}=&  - \bar{\gamma}\, {\mathbf m}\times \bigg[{{\mathbf H}_{eff}+\eta {\boldsymbol \Omega}{\mathbf m}}\bigg] - \lambda \, {\mathbf m}\times\bigg({\mathbf m}\times \bigg[{{\mathbf H}_{eff}}+\eta {\boldsymbol \Omega}{\mathbf m}\bigg]\bigg) .  
    	\end{align}      
 In the absence of material motion, i.e., $\bfOmega = \bfzero$, we refer to \eqref{eq:llg_simplified} as the `Landau-Lifshitz form of the LLG equation.' In the presence of material motion, the equation is an extension of the well-accepted LLG equation of micromagnetics to the regime of coupling with finite deformation of matter. Essentially, the effective magnetic field is found to be modified by contributions from the material spin tensor and elastic distortion of the continuum. 
 
 The original Landau-Lifshitz equation \citep{Landau1935} corresponds to \eqref{eq:llg_simplified} with $\bfOmega = \bfzero$ and $\xi = 0$ with $\bar{\gamma}, \lambda$ considered as independently variable parameters (and independent of $\xi$). Thus, \eqref{eq:llg_simplified}, which is a restatement of \eqref{eq:llg_direct}, shows that the original Landau-Lifshitz equation and the Gilbert form of the LLG equation (\eqref{eq:llg_direct} with $\bfOmega = \bfzero$) may be considered essentially as time-rescaled versions of each other.

\subsection{A dynamic model of magnetomechanics for $\psi = \psi(\bfW, \bfm)$}
In closing this section, we note that with a free energy density of the form $\psi = \psi(\bfW, \bfm)$, the resulting model developed in this Sec.~\ref{sec:theory} includes a dynamic extension of the constrained theory of magneto-striction/elasticity of \cite{james1993theory, desimone2002constrained}. Such an extension follows from the balance laws of mass, momentum, angular momentum, Maxwell's magnetostatics, and constrained by frame-indifference and (a mechanical version of) the Second Law, while allowing for dissipative effects of material viscosity, as well as damping due to the evolution of the magnetization relative to the material. The operative `constraint' of mechanical strain following that defined by the easy axis of magnetization is imposed in the DeSimone-James-Kinderlehrer model by an energetic penalty through a free-energy density that fits the type being discussed here; thus, no kinematic constraint, as understood in (continuum) mechanics, is enforced (in Sec.~\ref{sec:zhao_model} the kinematic constraint \eqref{eq:mevol_constr} is imposed to demonstrate a different constrained magnetomechanical model fully consistent with the principles of continuum mechanics). 

In such a dynamic extension, assuming $\partial_{\bfW} \p_{\bfm} \psi \neq \bfzero$, the evolution of the magnetization is coupled to motion through $\bfH_{eff}$ and through the dissipative coupling to the material spin in the balance of angular momentum \eqref{eq:llg_simplified}, and the material motion is coupled to magnetization through the equilibrium and dynamic stresses $\bfT = \bfT^e + \bfT^d$ in the balance of linear momentum \eqref{eq:balance_linear_momentum}. The work of \cite{benevsova2025variational} (and other prior works mentioned therein; of particular relevance here are their Remarks 2.3 and 2.5) also provides a different extension where angular momentum balance is not considered and neither is the Second Law in a direct manner (e.g., it is not clear whether their Cauchy stress tensor is symmetric or not, and if the physics of the model can be phrased without reference to a distinguished reference configuration embedded in three-dimensional space). 

It is perhaps worth noting here a somewhat obvious fact related to the absence of any non-local energy term as a function of magnetization in the expression of our free energy density, in contrast to the variational formulations in, e.g., \citep{james1993theory, desimone2002constrained, desimone2002reduced}. In a purely variational static setting the \emph{potential} energy is the main ingredient whose minimimization  as a functional of $(\bfx,\bfm)$ determines (appropriately defined) solutions to the problem. That energy necessarily contains the potential energy of the magnetic field working on the magnetization (cf.~\cite[Eqn.~19]{Landau1935}, \cite[Eqn.~2.10]{desimone2002constrained}), acting as a `loading' term. Correspondingly, in our PDE setting the `loading' due to the magnetic field (including the stray field) appears in the body torque term $\bfK$ in the balance of angular momentum through $\bfK_m$ \eqref{eq:K_m}. This follows directly from the fundamental physical meaning of how the magnetic induction induces a torque on the magnetization (or a magnetic dipole at the microscopic level) \cite[Eqn.~5.71, 5.1]{jacksonCED}, a fact that is intrinsically built into the recovery of the LLG equation within our setting. A difference between the variational/energy and PDE/balance-laws settings is that the latter fundamentally involves quantities like the stress, couple stress, body torques and body forces which have no immediate representation in the former, especially when dealt with the direct method of the calculus of variations. 

\section{A power-less rotational inertia term in the continuum mechanics of polar materials with classical kinetic energy density}\label{sec:add_rot_inertia}
We consider a polar material, with a director field $\mathbf m$, which supports couple stresses whose surface torques do work on the director spin vector field and which supports a general asymmetric Cauchy stress tensor field. Furthermore, we assume that its rate of change of angular momentum density is given by the expression
 \begin{align}\label{eq:def_a}
 \rho \, c_1 \, \dot{\mathbf m}+\dot {\mathbf a},
 \end{align}
 where $\mathbf a$ is a vector field which satisfies
 \begin{align}\label{eq:def_A}
 \dot{{\mathbf a}} = {\boldsymbol \omega}\times \rho {\mathcal A},
 \end{align}
 and ${\mathcal A}$ is another vector field, arbitrary up to having physical dimensions of $[\calA] = \frac{{Length}^2}{Time}$. For $c_1 = \frac{M_s}{\gamma}$ (see Sec.~\ref{sec:EDHB}) and $\calA = \bfzero$ we recover the micromagnetic theory considered so far. The vector field $\calA$ may be considered a director like state variable of `microscopic' origin whose spin contributes to the macroscopic angular momentum\footnote{ It seems not beyond the realm of possibility that this field and its induced torque may find utility in providing a placeholder for incorporating `spintronic' effects \citep{chen2023spintronic} within continuum mechanics, particularly facilitated by the lack of constitutive restrictions on it dictated by the macroscopic laws of continuum mechanics, as is subsequently shown herein.}; but at any rate, at this stage we are unable to make a definite physical statement about its nature beyond the above attributes. %%According to \cite{desimone1995inertial}, 
 We note here that J. Serrin \citep{serrin1991equations} advocated the introduction of powerless inertial terms in the mechanical balances. Furthermore, \cite{desimone1995inertial} introduced `Coriolis inertial forces' that appear in their balance laws but do not expend power and neither appear in the dissipation inequality.
 
 The material also satisfies balance of mass \eqref{eq:bom}, balance of linear momentum \eqref{eq:balance_linear_momentum}, and balance of angular momentum \eqref{eq:angular_momentum_0} modified to
 \begin{align}\label{eq:ang_mom_ref}
	       \rho c_1 \dot  {\mathbf m} + \dot{\mathbf a} &=div {\boldsymbol \Lambda} -{\mathbf X}:{\mathbf T}+\rho {\mathbf K}.
\end{align}

Examination of the power expended for such a polar material, in direct analogy with \eqref{eq:thm_power_expend}:
\begin{align}\label{eq:TPE_ref}
	{\mathcal P}_{ext}(t)&=\int_{{\mathcal V}(t)}\rho {\mathbf b}\cdot {\mathbf v} dv+\int_{\partial {\mathcal V}(t)}{\mathbf T} {\mathbf n}\cdot {\mathbf v}da+
	\int_{\partial {\mathcal V}(t)}{\boldsymbol  \Lambda} {\mathbf n}\cdot {\boldsymbol \omega}da+\int_{{\mathcal V}(t)}\rho {\mathbf K}\cdot {\boldsymbol \omega}dv \nonumber \\
	&=\int_{{\mathcal V}(t)}\big(\rho {\mathbf v}\cdot \dot{\mathbf v}
	+ 
	\rho ( c_1 \, \dot{\mathbf m} + {\boldsymbol{\omega}} \times {\mathcal A} ) \cdot {\boldsymbol \omega}  \big) 
	dv
	+\int_{{\mathcal V}(t)}\big({\mathbf T}:{\mathbf L}+{\boldsymbol \Lambda}:{\boldsymbol \Theta}+{\mathbf T}:{\boldsymbol \Gamma}\big) dv \nonumber \\
		&=\int_{{\mathcal V}(t)}\rho {\mathbf v}\cdot \dot{\mathbf v} dv
	+\int_{{\mathcal V}(t)}\big({\mathbf T}:{\mathbf L}+{\boldsymbol \Lambda}:{\boldsymbol \Theta}+{\mathbf T}:{\boldsymbol \Gamma}\big) dv,
\end{align}
reveals that if the rate of change of angular momentum can be assumed to be of the form \eqref{eq:def_a}-\eqref{eq:def_A}, then the power of the translational and rotational inertial forces retains its classical form and meaning, i.e.,
 \begin{equation*}
 \begin{aligned}
    & \mbox{Power of the (translational + rotational) inertial forces} = \\
    & \qquad \int_{\mathcal{V}(t)} \left( {\boldsymbol \omega} \cdot \rho \left( {\boldsymbol \omega} \times \left( c_1 {\mathbf m} +  {\mathcal A} \right)\right) + \rho \dot{{\mathbf v}} \cdot{ \mathbf v} \right) \, dv = \frac{d}{dt} \int_{\mathcal{V}(t)} \frac{1}{2}\rho |{\mathbf v}|^2 \, dv.
 \end{aligned}    
 \end{equation*}
and the same conclusion holds for the dissipation \eqref{eq:diss}.

 \emph{The implication of the above result is that the field $\mathcal A$ is completely free of constitutive constraints from frame-indifference and continuum thermodynamics, subject only to the balance laws (and corresponding boundary conditions) and modeling constraints arising from micro/macroscopic physical observations and/or microscopic physical theory.}

\subsection{A constrained theory of polar materials with materially-constrained director motion}\label{sec:constr_polar}
 A further observation is worthy of note. Suppose one wishes to consider a constrained polar material where the director spin vector field  
is constrained to be the material spin vector:
\begin{equation}\label{eq:constr_polar}
\mbox{a constrained polar material:} \qquad  {\boldsymbol \omega} = {\mathbf s} \qquad \mbox{so that} \qquad {\boldsymbol \Omega} = - {\boldsymbol \Gamma}.
\end{equation}
Thus
\begin{equation}\label{eq:mevol_constr}
\dot{\mathbf m} = {\mathbf s} \times {\mathbf m} = {\boldsymbol \Omega} {\mathbf m}.
\end{equation}
Then the external power supplied takes the form
 \begin{equation}\label{eq: Tskw_indep}
     {\mathcal P}_{ext}(t) = \int_{{\mathcal V}(t)}\rho {\mathbf v}\cdot \dot{\mathbf v} dv
	+\int_{{\mathcal V}(t)}\big({\mathbf T}_{sym}:{\mathbf D}+{\boldsymbol \Lambda}:{\boldsymbol \Theta}\big) dv,
 \end{equation}
with the dissipation given by
\begin{align}
	\mathcal D:={\mathcal P}_{ext}-\frac{d}{dt}\bigg[{\mathcal K}+{\mathcal F}\bigg]=\int_{{\mathcal V}(t)}\big({\mathbf T}_{sym}:{\mathbf D}+{\boldsymbol \Lambda}_{dev}:(\nabla \bfs)_{dev} -\rho \dot \psi\big) dv,
\end{align}
 noting that $\nabla \bfs = (\nabla \bfs)_{dev}$.

\emph{Thus, for such a  constrained polar material, the skew-symmetric part of the Cauchy stress tensor ${\mathbf T}_{skw}$ and the hydrostatic part of the couple stress tensor, $\frac{1}{3} tr(\bfLambda)$, are subject only to the balance laws and modeling constraints arising from micro/macroscopic physical observations and/or microscopic physical theory, without any constitutive restrictions from continuum thermodynamics.}\footnote{This observation is essentially due to R.~D.~Mindlin as described in footnote 2 of \cite{toupin1962elastic}, also see \cite{mindlin1962effects}.}

The thermodynamic relations follow from
\begin{equation}\label{eq:thermo_relation_constr}
\begin{aligned}
	{\mathcal D} & = \bigg[{\mathbf T}_{sym} + \bigg(\rho{\nabla {\mathbf m} }^{T}\frac{\partial \psi }{\partial \nabla {\mathbf m} }+
	\rho{\mathbf W}^{T}\frac{\partial \psi }{\partial {\mathbf W} }\bigg)_{sym}\bigg]:{\mathbf D} \\
	& \quad +\bigg[ \rho{\nabla {\mathbf m} }^{T}\frac{\partial \psi }{\partial \nabla {\mathbf m} } + \rho{\mathbf W}^{T}\frac{\partial \psi }{\partial {\mathbf W} } - \rho\frac{\partial \psi }{\partial {\mathbf m} }\otimes {\mathbf m} - \rho\frac{\partial \psi }{\partial \nabla {\mathbf m} } {\nabla {\mathbf m} }^{T}\bigg]_{skw}:{\boldsymbol \Omega} \\
	& \quad +\bigg[{\boldsymbol \Lambda} - \rho{\mathbf X}:{\mathbf m}\otimes\frac{\partial \psi }{\partial \nabla {\mathbf m} }\bigg]_{dev}:(\nabla \bfs)_{dev}\\
    & \quad + {\mathbf T}^d_{sym} : {\mathbf D} \geq 0.
\end{aligned}
\end{equation}
The second line vanishes on the use of a frame-indifferent free energy density obeying the Ericksen identity \eqref{eq:Ericksen_identity}. The constitutively determined equilibrium stress response is given by \eqref{eq:total_stress} with \emph{right-hand-side symmetrized}, and the couple stress by \eqref{eq:couple_stress}, interpreted as a relationship for only its deviatoric part.

\subsection{Example 1: Allowance for the Einstein-de Haas and Barnett effects within continuum mechanics}\label{sec:EDHB}
 \cite{einstein1915experimental} (preceded by \cite{richardson}) showed that a change in magnetization of a cylinder, oriented along a magnetic field-producing solenoid's axis, generates a rigid rotation of the body. \cite{barnett1915magnetization} showed the development of magnetization in a body due to applied rotation. 
 
 Below we provide preliminary arguments for the possibility of obtaining such effects within the model proposed in this work, one which does not affect the definition of the standard kinetic energy density of the body. 
 
Allowing for the additional rotational inertia term $\dot{\mathbf a}$ introduced in Sec.~\ref{sec:add_rot_inertia}, the rate of change of angular momentum density in the absence of the first two terms in $\boldsymbol{H}_{eff}$  \eqref{eq:_Heff} is given by (cf.~\eqref{eq:llg_simplest})
 \begin{align}\label{eq:Edh_1}
     {\boldsymbol \omega} \times \left( \rho\frac{M_{s}}{\gamma} {\mathbf m} + \rho {\mathcal A} \right) &= {\mathbf m}\times \left[\rho{ \mu_0}{\mathbf H}  - \beta_m \, \dot {\mathbf m} + \beta_m \, {\mathbf s} \times {\mathbf m}\right].
 \end{align}
 Introducing the constants
 \[
 c_1 = \frac{M_s}{\boldsymbol \gamma}; \qquad c_2 = \frac{\beta_m}{\rho}
 \]
and using the relation $\bfomega = \bfm \times \dot{\bfm}$, \eqref{eq:Edh_1} can be expressed as 
\begin{equation}\label{eq:edh_barnett_demo}
\begin{aligned}
        \mathbb{M}_A \,\dot{\mathbf m}  & = { \mu_0}\bfm \times \bfH + c_2 \, \bfs_{\perp \bfm}\\
& \\
         \mbox{where} \qquad \mathbb{M}_A & := \big[ g(\calA, \bfm) \bfI - c_2 {\mathbf X} {\mathbf m} - {\mathbf m} \otimes {\mathcal A} \big]\\
         g(\calA, \bfm) & := c_1 + \calA \cdot \bfm \\
         \bfs_{\perp \bfm} &:= \bfs - (\bfs \cdot \bfm) \bfm
\end{aligned}
\end{equation}
and we recall that $\bfX \bfm = \epsilon_{ijk} m_k \bfe_i \otimes \bfe_j$.

%%Before proceeding to explicit computations, 
A few general observations can now be made. For simplicity, we consider situations when the magnetic field is uniform in a body and the magnetization divergence-free within it (for a long cylinder, the stray field would then be uniform in the cylinder except perhaps near its ends). It is well understood that in the LLG theory, in the absence of damping ($c_2 = 0$), the LL field torque $\bfm \times \bfH$ cannot produce alignment of the magnetization with the magnetic field direction, causing only precession of the magnetization vector pointwise. The Gilbert damping term $c_2 \neq 0$ produces alignment. The classical LLG theory has no coupling to the material spin field. 

 For simplicity, let us assume a spatially uniform material spin field arising from rigid rotation and aligned (anti-)parallel to the magnetic field direction.  Let us also assume $\calA$ to be a function of $(\bfs, \bfm)$. Then, even in the absence of damping ($c_2 = 0$, $\mathbb{M}_A^{-1} = (g(\calA, \bfm))^{-1} \bfI + \bfm \otimes \calA$ for $\calA \perp \bfm$, e.g., $\calA = f(\bfs \cdot \bfm) \bfs \times \bfm$ where $f$ is a scalar valued function (say, e.g., a constant or signum) of its argument, if the initial condition on $\bfm$ does not everywhere (along the whole cylinder) lie on the plane perpendicular to the magnetic/material spin axis, then there can be alignment of the magnetization distribution with the (material) spin axis. Given the vast freedom in the possibilities for $\calA$, e.g., the explicit form mentioned above, it can be checked that anti-alignment of the volume averaged magnetization in the body with the spin axis is possible, starting from a distribution with no \emph{net} magnetization in the body. Of course, non-vanishing damping promotes alignment without any restriction on the initial magnetization distribution. These features would be in accord with the Barnett effect \citep{barnett1915magnetization}, which corresponds to inducing magnetization through material spin. In fact, even in the absence of an external field or dissipation and with the initial magnetization lying in the plane perpendicular to the material spin ($\bfm \perp \bfs$ along the whole cylinder), a tendency towards alignment can be obtained by a suitable choice of $\calA$; e.g., for $\hat{\bfs}$ the unit vector in the direction of $\bfs$, consider $\calA := - c_1 \bfm + b_1 \hat{\bfs} \times \bfm$, $b_1$ a scalar.

As for the E-dH effect, assuming  any time-dependent spatially uniform material spin field as above (for simplicity), the magnetization in the body can be solved for from \eqref{eq:edh_barnett_demo} given an initial condition on $\bfm$, with and without damping. Any such evolving magnetization with net magnetization change in the body can be interpreted as inducing a mechanical spin. Moreover, it is conceivable to think of magnetizations that produce no net torque in the body even from the term involving the magnetic field - e.g., axially symmetric magnetizations in a cylinder. %%We demonstrate some such cases below in Sec.~\ref{sec:compute_result}. 
The material spin fields involved, when conceived of in a cylinder, require radial mechanical forces in the body to sustain them. In a rigid body approximation, this is provided by the constraint of rigidity. In a deformable body, this would require radial stresses whose divergence would balance the inertia of axisymmetric spinning. Without external boundary tractions, such radial stress fields would be radially inhomogenous. In case boundary conditions support them, it is most likely that the boundary traction would be radially directed and produce no torque about the cylinder axis.

%%{ to be continued - section with computed results.}
 
 \subsection{Example 2: Constrained magnetization dynamics of hard magnetic soft materials~\citep{zhao2019mechanics}}\label{sec:zhao_model}
 Recently,~\citet{zhao2019mechanics} illustrated a nonlinear field theory for hard magnetic soft materials under magnetic fields. In their model, the magnetic particles in an idealized hard magnetic soft material magnetically saturate when exposed to a strong magnetizing field. With ${\mathbfcal M}^{0}$ $\left([\calM_0]= \frac{Charge.Length^2}{Time.Mass}\right)$  to be considered as the remnant magnetization, they define 
the non-symmetric Cauchy stress as 
\begin{equation}
    {\mathbf T}=\frac{1}{J}\frac{\partial \psi}{\partial {\mathbf F}}{{\mathbf F}}^{T}-\rho {\mathbf H}_{ext}\otimes {\mathbfcal M}^{0}.
    \label{eq:cauchystress}
\end{equation}
A motivation of our work is to try to understand the theoretical structure of this demonstrably successful nonlinear model of micromagnetics for hard magnetic materials (\cite{zhao2019mechanics, wang2021evolutionary}) that entails the claim that ``rigid body rotation can change the free-energies of polar materials as well.'' 

\emph{We show here how the model employed by these authors can be recovered as a fully frame-indifferent theory for a constrained micromagnetic material whose free-energy density is manifestly invariant to rigid body rotations.}

We first note that, as postulated, \eqref{eq:cauchystress} does not have the correct invariance for a tensor on the current configuration and the remnant magnetization should be pushed forward to the current configuration (also see \citep{dorfmann2024hard} on this point).

In any case, our goal is to consider the material considered by Zhao and co-workers as a constrained polar material in the sense of \eqref{eq:constr_polar}-\eqref{eq:mevol_constr} and furthermore make the choice
\begin{equation}\label{eq:A_neutralize_angmom}
    {\mathcal A} = - c_1 {\mathbf m};
\end{equation}
such a choice may be interpreted as setting up the power-free constraint torque ($\bfomega \times \calA$) to maintain the kinematic constraint given by \eqref{eq:A_neutralize_angmom}. In that case, balance of angular momentum for the model reads as
 \begin{align}\label{eq:ang_mom_ref_1}
	     {\mathbf 0} &=div {\boldsymbol \Lambda} -{\mathbf X}:{\mathbf T}+\rho {\mathbf K}.
\end{align}
The net effect of the assumption \eqref{eq:A_neutralize_angmom} is along the lines of what some authors think of the situation appropriate for nematic liquid crystals; to quote \cite{leslie1992continuum} ``The inertial term
associated with local rotation of the material element is omitted because in
general it is negligible.''
Furthermore, we consider a free energy density that only depends on ${\mathbf F} = {\mathbf W}^{-1}$, and thermodynamic consistency of the model follows from Sec.~\ref{sec:constr_polar}; of particular importance is the result from Sec.~\ref{sec:constr_polar} that the skew-symmetric part of the stress tensor is constitutively undetermined, to be determined from the field equations and boundary and initial conditions\footnote{Much like the pressure field when the constraint of incompressibility is in force.} as, indeed, what happens in \eqref{eq:tskw_zhao} below. Then, from \eqref{eq:thermo_relation_constr} and \eqref{eq:couple_stress}, the couple stress tensor vanishes identically and, assuming no dissipative stress contribution for the material, the equilibrium stress is given by
\begin{equation}\label{eq:tsym_zhao}
	{\mathbf T}_{sym} = - \rho{\mathbf W}^{T}\frac{\partial \psi }{\partial {\mathbf W} } = \frac{\rho_0}{J}\frac{\partial \psi}{\partial {\mathbf F}}{\mathbf F}^T
\end{equation}
when balance of mass \eqref{eq:bom} is satisfied in the alternate form $\rho = \frac{\rho_0}{J}$, where $\rho_0$ is the mass density field on a reference configuration and $J = \det {\mathbf F}$, where ${\mathbf F}$ is the deformation gradient from that reference.  Furthermore, the Ericksen identity \eqref{eq:Ericksen_identity} guarantees that for any frame-indifferent free energy density functional form being considered here (i.e., ${\psi} = {\psi}({\mathbf F})$), the right-hand-side of \eqref{eq:tsym_zhao} is symmetric even without explicit symmetrization.

The micromagnetic body torque is given by
\[
\rho {\mathbf K} = \rho { \mu_0} \,  \calM_s {\mathbf m} \times {\mathbf H},
\]
and solving balance of angular momentum {\color{teal} \eqref{eq:ang_mom_ref}/\eqref{eq:ang_mom_ref_1}} then gives
\begin{equation}\label{eq:tskw_zhao}
    {\mathbf T}_{skw} = \rho { \mu_0} \left( \calM_s{\mathbf m} \otimes {\mathbf H}\right)_{skw} = - \rho { \mu_0} \left({\mathbf H} \otimes \calM_s {\mathbf m}\right)_{skw}.
\end{equation}
To summarize, there are two operative kinematic constraints in the model, \eqref{eq:constr_polar}-\eqref{eq:mevol_constr} and \eqref{eq:A_neutralize_angmom}, with corresponding constraint `forces' given by $\bfomega \times \calA$ and \eqref{eq:tskw_zhao}, respectively.

Putting \eqref{eq:tsym_zhao} and \eqref{eq:tskw_zhao} together, we recover the stress response used by \cite{zhao2019mechanics} to solve linear momentum balance in \citep{zhao2019mechanics,wang2021evolutionary}  with \eqref{eq:cauchystress} (modified for proper rotational invariance and with $\bfH_{ext}$ replaced by $\bfH = \bfH_a + \bfH_s$) as the equation for the stress. 

The derivation above justifies this assumption as an embodiment of solving angular momentum balance along with satisfying the thermodynamic and frame-indifference related restrictions of a constrained theory of micromagnetics. Moreover, it provides a natural extension of the constrained model of \cite{zhao2019mechanics} to incorporate couple stress effects due to magnetization gradients, and dissipative effects due to material viscoelasticity.

\section{Observations on treating the Second Law as a field equation and solution/approximation schemes for the proposed models}\label{sec:dual}
When the expression for the external power supplied to the body is considered to exclude the torque density arising from the demagnetization field and hence given by $\bfK - (\mu_0 \calM_s) \bfm \times \bfH_s$, $\calP_{ext}$ \eqref{eq:red_pwr_exp} takes the form
\begin{equation*}
    {\mathcal P}_{ext}(t) = \int_{{\mathcal V}(t)}\rho {\mathbf v}\cdot \dot{\mathbf v} dv
	+\int_{{\mathcal V}(t)}\Big({\mathbf T}:{\mathbf L}+{\boldsymbol \Lambda}:{\boldsymbol \Theta}+{\mathbf T}:{\boldsymbol \Gamma} - \bfomega \cdot \big(\rho \mu_0 \calM_s \,\bfm \times \bfH_s\big) \Big) dv,
\end{equation*}
resulting in the local form of the Second law (for the purposes herein) to be
\begin{equation}
    \label{eq:diss_H_s}
    {\mathbf T}:{\mathbf L}+{\boldsymbol \Lambda}:{\boldsymbol \Theta}+\big( {\mathbf T} - \rho \mu_0 \calM_s \, \bfm \otimes \bfH_s \big):{\boldsymbol \Gamma} - \rho \dot{\psi}  \geq 0.
\end{equation}
We note that  such a choice of $\calP_{ext}$ does not affect the body couple $\bfK$ appearing in the balance of angular momentum which includes $\bfK_m$ \eqref{eq:K_m}.

We now consider a point of view in which constitutive equations are specified for \emph{parts} of $\bfT, \bfLambda, \psi$ based on the best physical knowledge available, and the Second Law in the form \eqref{eq:diss_H_s} is required to be satisfied as a governing constraint, exactly like the other balance laws of mass, momentum, angular momentum, and the equations of magnetostatics\footnote{ It is our belief that such an approach can be beneficial when considering the full coupling of continuum mechanics with Maxwell's electrodynamics.}. Thus we write 
\begin{equation*}
    \begin{aligned}
        \bfT & = \bfT^c + \bfT^u\\
        \bfLambda & = \bfLambda^c + \bfLambda^u\\
        \psi & = \psi^c + \psi^u
    \end{aligned}
\end{equation*} 
where the parts $(\cdot)^c$ are constitutively specified, and the parts $(\cdot)^u$ are left as undetermined fields at this point. Such a modeling choice reflects the inevitable lack of full knowledge of material behavior and the governing rules for the dynamical processes that continuum bodies undergo.

Consider now the balance of angular momentum written in the form
\begin{equation}\label{eq:bam}
    \rho \frac{M_s}{\gamma} \dot  {\mathbf m} + (\bfm \times \dot{\bfm}) \times {\mathcal A} =div {\boldsymbol \Lambda} -{\mathbf X}:{\mathbf T}+\rho {\mathbf K},
\end{equation}
\begin{comment}
For specificity in the discussion, let us assume, for example, that the `best' constitutive assumptions that can be deployed are consistent with the relations \eqref{eq:total_stress},\eqref{eq:couple_stress}, and \eqref{eq:viscous_stress} (arising from the assumption that $\calP_{ext}$ is given by \eqref{eq:pwr_exp}). As already observed, $\calA$ is left unconstrained even by the minimal requirements on constitutive response that arise from thermodynamics. However, its inclusion appears to allow the accommodation of observed physical effects within a model of continuum mechanics, as discussed in Sec.~\ref{sec:EDHB}.
\end{comment}
and also \eqref{eq:diss_H_s} stated in the form of an equality,
\begin{equation}
    \label{eq:diss_H_s_plus_s}
    {\mathbf T}:{\mathbf L}+{\boldsymbol \Lambda}:{\boldsymbol \Theta}+\big( {\mathbf T} - \rho \mu_0 \calM_s \, \bfm \otimes \bfH_s \big):{\boldsymbol \Gamma}  - \rho \dot{\psi} - s^2 = 0,
\end{equation}
where $s$ is a real-valued field defined on the body for all instants of time.

The problem to be solved is to determine a set of fields $(\rho, \bfv, \bfm, s, \bfH_s, \calA, \bfT^u, \bfLambda^u, \psi^u)$ that satisfy the balances of mass, momentum, angular momentum, the equations of magnetostatics\footnote{As already alluded to, the stray field $\bfH_s$ can be exchanged for a scalar potential in terms of solution variables, with the attendant elimination of \eqref{eq:mag_stat}$_1$ as a governing equation.}, and the second law in the form \eqref{eq:diss_H_s_plus_s}, given a set of constitutive equations for $(\bfT^c, \bfLambda^c, \psi^c)$. 

Suppose now that experimental measurements/microscopic modeling can provide some guidance on the nature of the field $\calA$ on the body as a function of space and time, say $\bar{\calA}$, in any particular problem (possibly as a constitutive statement) - if none is available $\bar{\calA} = \bfzero$ may be the considered choice, representative of the limited knowledge available. Similarly, in the spirit of invoking maximum dissipation, let a constant space-time field $\bar{s}$ be defined with $\bar{s}/\bar{s}_0 \gg 1$, where $\bar{s}_0 > 0$ is a scaling constant representing a typical magnitude of supplied power relevant to the problem (for non-dimensionalization). Then one can require that a functional of the following type be minimized:
\[
\int_0^T \int_{\calV(t)} a_1 |\calA - \bar{\calA}|^{p_1} + a_2 |s - \bar{s}|^{p2} + a_3 |\bfT^u|^{p_3} + a_4 |\bfLambda^u|^{p_4} + a_5 |\psi^u|^{p_5}\, dv dt,
\]
where $[0,T]$ is a given interval of time, $\{a_i|i = 1,\ldots,5\}$ are physically required dimensional constants and $\{p_i|i = 1,\ldots,5, p_i \geq 1\}$ such that each individual term in the integrand is a positive quantity of identical physical dimensions (typically that of energy density or $Energy/Length^3$ with the integral having dimension of $action$ given by $Energy. Time$). The minimization is carried out over a set each of whose elements is the list of space-time fields ($\rho$, $\bfv$, $\bfm$, $s$, $\bfH_s$, $\calA$, $\bfT^u$, $\bfLambda^u$, $\psi^u$), subject to the constraints of balances of mass  \eqref{eq:bom}, momentum \eqref{eq:balance_linear_momentum}, angular momentum \eqref{eq:bam}, magnetostatics \eqref{eq:mag_stat}, and the second law in the form \eqref{eq:diss_H_s_plus_s}. 

Roughly speaking, the extremality conditions for imposing such a constrained variational problem contains extra relations beyond the above constraints, such relations serving to constrain the added fields $(\calA, s,\bfT^u, \bfLambda^u, \psi^u)$\footnote{This is analogous to how control variables are constrained in the problems of Optimal Control governed by the Pontryagin Maximum Principle (cf.~\cite{evans_control, pontryagin}).}. Thus, the list $P := (a_i, p_i,\bar{s}, \calA)$ plays a role analogous to constitutive parameters in the model.

We note that any solution to the extremality condition of the above constrained variational problem will necessarily satisfy the constraint equations which are the ones that our continuum theory requires to be solved in as well-posed a manner as possible. Moreover, the minimization would in principle force the dissipation in the process to be as large as possible (since $\bar{s}$ is prescribed to be large), $\calA$ to be as close as possible to the experimental/best-guess-based-on-available-knowledge field $\bar{\calA}$, and $\bfT^u, \bfLambda^u, \psi^u$ to be as close as possible to vanishing, implying a degree of confidence in the assumed constitutive equations for the $(\cdot)^c$ parts of the stress, couple stress, and the free energy density. 

In such a formulation, the number of equations arising from the constraints of balances of mass, momentum, angular momentum, second law, and magnetostatics is less than the number of solution fields in the problem. Thus, if a variational principle could be set up to the above extremization problem in a number of fields that equals the number of constraint equations, and its Euler-Lagrange (E-L) equations impose (formally, at least) \emph{only the constraint equations and their initial and boundary conditions and nothing more}, that would be an ideal first step. In that case, a solution to the E-L system corresponding to a specific choice of the list $P$ would amount to selecting some solution out of the possible ones of the primal system (when solutions to the primal system exist, of course). In case a unique solution exists to the latter, any choice of the list $P$ that leads to a solution of the E-L system must recover the unique solution of the primal system. 

The above requirements on the design of a constrained variational problem (with reasonable mathematical properties, of course) are non-trivial, especially for the nonlinearities that typically arise from models of continuum mechanics and in treating the Second Law as a constraint equation to be solved on par with the other balances and not necessarily as a device to generate restrictions on constitutive equations \citep{ach_MRC}.  A scheme with such properties has recently been developed in \citep{ach2, ach3, AV_nash}, with close connection to the work of \cite{CMP18,Br20} on Hidden Convexity in nonlinear PDE, and Optimal Transport. Applications of the approach to various field theories and involving mathematical analysis are demonstrated in  \cite{V22,V25,ASZ24,AGS24,sga}. While still in its infancy, encouraging progress has also been demonstrated in approximating solutions of nonlinear differential equations related to continuum mechanics in  \cite{ka1,sga,ka2}, \cite[which also contains mathematical analysis]{kpa} using the approach. These demonstrations involve the nonlinear system of ODEs of Euler for the dynamics of a rigid body, nonconvex elasto\emph{dynamics} and statics of a bar (without higher gradient regularization), (inviscid) Burgers equation, and the problem of traveling waves of a dispersive, nonlocal, nonlinear semi-discrete Burgers equation. Furthermore, \cite{Ach6} show applications of the idea to the case of dissipative Newtonian particle mechanics with anholonomic constraints and \cite{AG_control} to problems of optimal control.

It can be hoped that such a solution approach to the models here can be gainfully employed to explore their predictive capability w.r.t magnetomechanical response - even in the case when $\calA \equiv \bfzero$ and $\calP_{ext}$ is given by \eqref{eq:pwr_exp}, with constitutive equations chosen in a manner to identically satisfy the Second Law, i.e., the theory developed in Sec.~\ref{sec:theory} of this work (in which case the fields $(\cdot)^u \equiv \bfzero$ can be imposed a-priori).

\section*{Acknowledgement}
AA acknowledges support from NSF OIA-DMR grant \# 2021019. SP greatly acknowledges funding support from 
Mathematical Research Impact Centric Support, Science and Engineering Research Board (MATRICS-SERB), India with grant no MTR/2020/000402. 

\appendix
\section{Appendix - Literature Review}\label{app:lit_rev}
%%
\begin{comment}
Utilizing the evolution of complex magnetic domain patterns under an external magnetic field and its strong coupling with mechanical fields is the fundamental principle behind the development of various functional solids and liquids for various engineering applications~\citep{sutrisno2015recent,huang2020recent,li2021magnetoresistive}. Primary examples of such functional materials are ferromagnetic or magneto-rheological materials (MRs) where ferromagnetic particles are dispersed in a solid or fluid. Depending on the type of medium, MRs are broadly classified as a magneto-rheological elastomer (MRE) or a magneto-rheological fluid (MRF)~\citep{carlson2000mr}. These materials can reversibly change their effective mechanical or rheological properties on the application of an external magnetic field. In general, a ferromagnetic material exhibits magneto-mechanical interaction at the micro-scale through magnetization reorientation and magnetic domain wall motions. Emergence of spontaneous deformation accompanying the spontaneous magnetization leads to the phenomenon of magnetostriction.
\end{comment}

 The seminal work of Landau and Lifshitz offered the mathematical foundation of the statics and dynamics of magnetization of rigid continua~\citep{Landau1935}. To consider dynamic micromagnetism, the Landau-Lifshitz-Gilbert (LLG) equation~\citep{gilbert2004phenomenological, gilbert1956formulation} is widely used. This equation describes the Larmor precession under an effective field, with an additional damping term that relates the magnetic term to the local environment. It successfully captures the evolution of magnetization in a specimen when subjected to alternating currents. 
 
To understand the macroscopic behavior of a non-conducting magnetically saturated media undergoing large deformation, ~\cite{tiersten1964coupled} developed governing equations and boundary conditions modeling the interaction between an electronic spin continuum and a lattice continuum. Later ~\cite{tiersten1965variational} presented a variational principle for the aforementioned nonlinear differential equation and boundary conditions in the absence of heat flow and dissipation. %%These theories suffer the difficulties arising from accounting for the interaction between point-like polarized and magnetized particles with the deforming continuum. 
Employing matter-on-matter interaction within a magnetized body, ~\citet{brown1966magnetoelastic} developed the static theory of micromagnetics where the coupling of the magnetization with the deformation field is considered in a static variational context. ~\cite{maugin1976continuum_a} introduced local and global field equations for finitely deformable ferromagnets and anti-ferromagnets from a phenomenological viewpoint. In a follow-up work  ~\cite{maugin1976continuum_b} employed thermodynamic principles in formulating the constitutive laws for internal forces for reversible processes in accordance with Coleman's thermodynamics and for irreversible processes with Onsager-Casimir theory.~\cite{desimone1996continuum} constructed a continuum theory of deformable micromagnetics considering two main features of mutual interaction (mutual forces and mutual torque) i.e., mechanical interaction among magnetized body parts as proposed by ~\cite{brown1966magnetoelastic},  and explicit use of microstructure as adopted by~\cite{tiersten1964coupled}. The theory of micromagnetics has been further generalized and rigorously analyzed to capture macroscopic behavior with high anisotropy~\citep{desimone2002constrained, desimone2002reduced}. We also mention here the recent work of \citet{benevsova2025variational} for  mathematically rigorous results on the existence of solutions to a model of magnetoelasticity with  dissipation developed by them, as well as a review of the more mathematically oriented literature.

Significant research efforts have also been carried out over the last decades in developing models for magnetorheological elastomers with desired functional behaviors~\citep{kankanala2004finitely,dorfmann2004nonlinear,danas2012experiments,haldar2016finite,romeis2017theoretical}. The developed MREs are often based on soft-magnetic materials such as iron and iron-nickel alloys having low coercivity which offer simple elongation and compression under external magnetic fields. To attain shape-programmability for MREs, ~\citet{zhao2019mechanics} incorporated magnetically hard-particles with high coercivity in a soft polymer and developed a nonlinear field theory to account for coupling of finite deformation and a magnetic field. Nevertheless, the research works on both soft-magnetic and hard-magnetic materials based MREs mostly employ the principles of the static theory of magnetization. To deal with magnetization dynamics of ferromagnetic particles embedded in a soft solid, ~\citet{keip2019variationally} proposed a finite deformation micromagnetically informed continuum framework. Mainly considering 
the magnetization as a phase field variable, an evolution equation for the magnetization coupled to large deformation is proposed, which is essentially consistent with Landau-Lifshitz equation of micromagnetics without precessional term. This theory also recognizes the schemes proposed by ~\citet{miehe2012geometrically,sridhar2016homogenization} for ferromagnetic materials when small deformation kinematics is assumed.

\newpage
\bibliographystyle{apalike}
\bibliography{reference,nash} 
%------------all references---------------------------------------------------
\end{document}